# Quantum Darwinism and Computability Theory

Subhash Kak[1]
October 23, 2014

**Abstract**. This paper examines whether unitary evolution alone is sufficient to explain emergence of the classical world from the perspective of computability theory. Specifically, it looks at the problem of how the choice related to the measurement is made by the observer viewed as a quantum system. In interpretations where the system together with the observers is completely described by unitary transformations, the observer cannot make any choices and so measurement is impossible. From the perspective of computability theory, a quantum machine cannot halt and so it cannot observe the "computed" state, indicating that unitarity alone does not explain all matter processes. Further it is argued that the consideration of information and observation requires an overarching system of knowledge and expectations about outcomes.

**Introduction**
Zurek presents the idea of quantum Darwinism to explain the emergence of the classical world based on unitarity alone [1]. He claims that this solves the problem why classical states are robust and remain unperturbed by measurement. The basic idea is to see measurement as an interaction between the system $S$ and its environment $E$ that leads to the singling out of the eigenvectors (pointer states) of the environment in such a way that the phase relations between the pointer states are lost [2]. In this interaction, the state of the system is represented by the reduced density matrix $\rho_S$ which is obtained from the composite state $\Psi_{SE}$ of $S$ and $E$ by tracing out the environment: $\rho_S = Tr_E |\Psi_{SE}\rangle\langle\Psi_{SE}|$. The loss of the phase information leads to classical or probabilistically additive behavior.

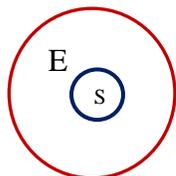

Figure 1. Universe viewed as System (S) and Environment (E)

But in this view there is nothing that distinguishes $S$ from $E$ and therefore it implicitly leaves out the special role of observer in the measurement process. Since classical observations are repeatable, Zurek brings in the classicality of the observation process indirectly by assuming that correlations between fragments F of the environment E and the system are considered to account for the repeatability associated with the classical states. In quantum Darwinism, these fragments are taken to be independent and, therefore, each of these fragments as an observer will get the same information about the system, leading to classical repeatability.

The idea of a computation determined by the environment is not, in itself, surprising. Consider ants in random motion on a mountainside so that some are able to reach the mountaintop. The fact of having reached the top cannot be ascribed to the intention of the ant, but rather to the nature of the environment. It is in that sense that the environment reduces the dynamics of the system. But if the environment itself is random, no such discernible "computation" can occur.

[1] Oklahoma State University, Stillwater, OK 74078

Zurek asserts that the mutual information *I* between the system *S* and the fragment *F* of the environment, available through decoherence, will satisfy:

$$I(S:F) = H_S + H_F - H_{S,F} \tag{1}$$

where the H functions represent the corresponding entropy values. He claims that states of *SE* contain almost all (all but δ) information $H_S$ about *S* in small fragments $f_\delta$ of *E*. This implies a redundancy of $R_\delta = 1/f_\delta$ in obtaining (1-δ) of information from the very many fragments. He concludes that "Large redundancy implies objectivity: The state of the system can be found out indirectly and independently by many observers, who will agree about their conclusions. Thus *Quantum Darwinism accounts for the emergence of objective existence*." [1]

This argument contradicts the notion of one universe with its wavefunction that evolves as a single system that is implicit in unitarity alone dynamics assumed by Zurek. It needs the postulate that subenvironments of the environment are independent (redundancy paradigm of Figure 1) and there exist independent observers (who constitute collections of subenvironments) both of which are contrary to the idea of a single wavefunction.

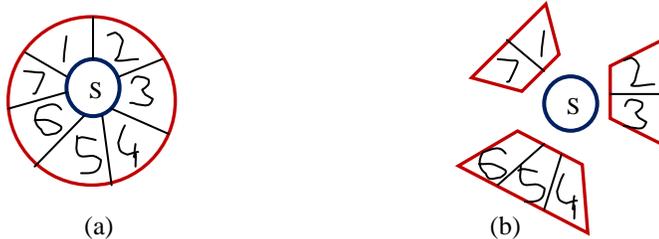

(a)          (b)
Figure 2. (a) The environment E divided into 7 subenvironments; (b) fragments are aggregations of subenvironments and each has nearly complete information (after [1])

Kastner [2] criticizes the idea that Zurek's "einselection" (environmentally induced superselection) can explain stable pointer states and she claims that it suffers from circularity of logic. According to her:

> "[The] problems is not so much a lack of observer-independence as it is a *failure to account for the initial independence of the environment from the system it is measuring*. That is, even if it is true that the system's only correlation with the environment is via the interaction Hamiltonian, and the environmental systems are randomly phased with respect to each other, these conditions cannot be explained from within the Everettian account: in that account, random phases are fictions. And these conditions are crucial to 'deriving' decoherence and the appearance of classicality in the no-collapse, unitary only Everettian picture. Thus, in that picture, apparently *de facto* classicality is crucial to deriving classicality." [2]

Kastner's critique is compelling. Others have likewise criticized the claim that unitarity alone is sufficient to explain measurement. The very definition of subsystems comes with ambiguity since the partition can be done in a variety of ways that are dependent on what the observer is interested in [3]. Fields argues that the idea of quantum Darwinism requires an extra-theoretical assumption [4].

Clearly, the multiplicities of views on the measurement problem arise from the differing interpretations of the quantum formalism. Measurement seen as decoherence [5]-[8] has the virtue of consistency but it leads to a variety of paradoxes like the information paradox [9] and as Kastner argues it cannot explain how classicality arises.

Nevertheless the question of quantum Darwinism, where the system together with the environment evolves into preferred states remains interesting. If unitarity alone is insufficient, what else might be needed for this to be possible? In order to advance this discussion further we reexamine the ideas of information and of observers that are fundamental to a consideration of the measurement problem from the perspective of quantum computability theory.

**Observer and information**
In extension of our self-understanding as biological organisms, an observer is a physical system but he or she is not described exclusively as a physical system for there may be other states such as emergent states that should be counted. To the extent that machines that are used to make observations are physical systems, they are extensions of human agents. The agent is further associated with internal states and a cultural context. Although some physicists believe that all phenomena must be reducible to physics, others have suggested that one needs different kinds of descriptions. Popper and Eccles believed one needs three different worlds, each with its own language, to understand reality [10]. In their classification, World 1 is the world of physical objects and states, World 2 is the world of subjective states such as thoughts, memories, and emotional states, and World 3 is knowledge in an objective sense. These worlds deal with outer sense, inner sense, and culture and spirit (or self), respectively. Eccles even argued that the self controls the brain [11].

In any event, information requires an observer for whom the prior expectations are changed once a message has been received. The central idea here is that of computation of global states and an underlying structure of data within which the message must he examined. An observer cannot analyze data without models and expectations on measurements. The observer, therefore, must not only belong to Eccles Worlds 1 and 2, but also World 3. The observer must have complex stable structures that are entangled not only within its subsystems but also with the environment. This implies that the assumption that the environmental systems are randomly phased as assumed by Zurek is not valid. What is true of the external environment is also valid of the internal environment (structure) of the observer. The claim can be made that whatever is observed is based on the cognitive systems that have the capacity for such observation. One may also take particular characteristics of reality to be a consequence of the nature of the cognitive system of the human agent [12].

There are many ways to look at information. More commonly, we view scientific information or knowledge within the context of specific areas that allows us to understand relationships between variables, find laws, and make predictions, where some of the phenomena are deterministic and others are random. Another perspective is to consider the nature of the language in which knowledge is expressed. Different areas of science have their own specialized languages and across fields the languages are not always consistent. This inconsistency points to the gaps that exist in our understanding of reality.

Interpretations of quantum theory provide a bridge between the formalism and intuition. A valid question to ask is if an interpretation can be checked for consistency within itself and the larger philosophical framework of quantum theory. The orthodox interpretation acknowledges a split between the quantum mechanical process and its measurement: the evolution of the system is unitary and the measurement is non-unitary and this embodies the intuitions related to the nature of the physical and the psychological worlds. The unitarity alone idea encapsulates the philosophical view that ultimately all human behavior will be reducible to machine-like behavior.

Now consider a point of comparison of the above with the classical world. Note that even though the three-body has no general analytical solution, measurement is not normally a problem in the classical world because of the clear hierarchical distance that exists between the observer and the system on which the measurement is being made. In computer science, the observer sits beyond and above the system examined by him. Thus in Figure 3, classes 1, 2, and 3 may be based on the size of the system and an example of this is the nation (class 1), the city (class 2), and the human observer (class 3). Clearly what happens to a member of C has little influence on the dynamics of class 2, and so on.

Now if we anthropomorphize the members of the three classes in a different kind of a system, then A who sits above B can observe the latter, so long as it is isolated from other systems in its own level (such as C) and the interactions with systems in the lower levels (such as D, E, and F) are small enough to be taken as background noise. In computer science or in machine theory also, the observer or the controller is hierarchically situated above the system.

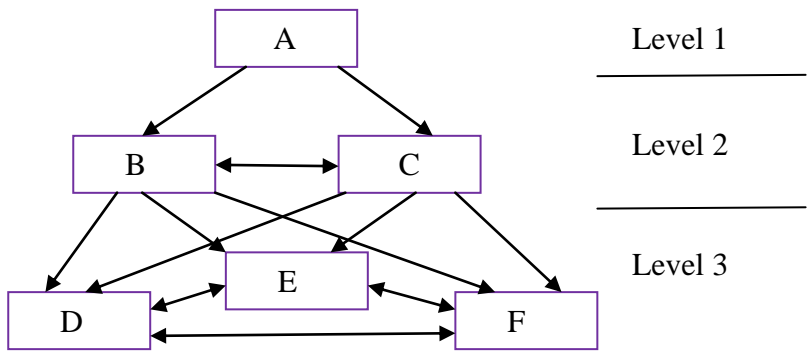

Figure 3. Universe with three hierarchical levels

From another perspective, a classical system can be observed since it can be isolated from other systems – and from the observer -- due to the local nature of interactions. In contrast, since interactions in quantum mechanics are nonlocal, it forebodes new problems in the very process of observation in a universe that is purely quantum mechanical.

The difference between the classical and the quantum modes has been investigated in many different ways (e.g. [13]) and it may be defined in terms of the measure of nonlocality [14]. At the quantum mechanical level, a system cannot be described in terms of clear hierarchies such as that of Figure 3. For example, an electron or a photon can be entangled with an atom [15].

**Computability in a tangled hierarchy**
In the absence of a clear hierarchy and in the presence of nonlocal interactions, subsystems cannot be isolated within a system. Such isolation is essential for the observer to recognize his or her own identity and this requires some knowledge of the environment in advance.

If systems cannot be completely separated before the observation begins, observation will be impossible. From another perspective, the need for global states (related to World 3 aspects of observation) in the observation process implies that the measurement of one system by another by unitarity alone cannot be reduced to the Turing machine formalism.

A quantum Turing machine (QTM) is an abstract machine, which enlarges the classical model of a Turing machine by allowing a quantum transition function. In a QTM, superpositions and interferences of configurations are allowed, but the controllers running the quantum algorithms and inputs and outputs of the machine remain classical [16],[17]; an example of this is quantum teleportation that requires sending some information on a classical channel. Without postulating a higher level measurement process (which cannot be quantum) the completion of the computation cannot be determined.

A purely quantum Turing machine cannot halt due to the continuing unfolding of its dynamics. From this perspective, unitarity alone cannot lead to measurement. In contrast, classical algorithms can converge to specific states as, for example, in the 3n+1 mapping [18]. Such converged states can be used as marker of the completion of a computation.

**Conclusions**
We would also like to draw a parallel of quantum Darwinism with the problem of measurement (or recognition) in neuroscience. It is no doubt that the incoming sensory signals are correlated with activity in specific regions of the brain just like the correlation between the system and the environment. But how this activity leads to measurement cannot be resolved within the framework of brain states since it leads to the homunculus problem.

Quantum Darwinism appears to embody the philosophy of random evolution even if one were to concede that "einselection" leads to the survival of the most proliferated states. Darwinism is supposed to encapsulate standard evolutionary framework in which new variation arises through random genetic mutation, inheritance occurs through DNA, and natural selection explains adaptation that makes organisms become well-suited to their environments. More recent research shows that "physical development influences the generation of variation (developmental bias); the environment directly shapes organisms' traits (plasticity); organisms modify environments (niche construction); and organisms transmit more than genes across generations (extra-genetic inheritance)." [19] This complexity in biological evolution appears to call for a corresponding complexity in our conception of observers that will be achieved only if we are prepared to concede not only processing of local information but also global states. Such global state processing is not possible in unitarity-alone models.

*Acknowledgements.* This research was supported in part by the National Science Foundation and by a grant from the Federico and Elvia Faggin Foundation.